\documentclass[]{aastex631}

\usepackage{xcolor}
\def\r{\textcolor{red}}
\def\b{\textcolor{blue}}
\def\p{\textcolor{purple}}
\usepackage{subcaption}
\usepackage{float}

\begin{document}

\title{The `blazar sequence' in TeV band}

\author{Zhihao Ouyang}
\affiliation{Shanghai Key Lab for Astrophysics, Shanghai Normal University, Shanghai, 200234, China}

\author[0000-0001-8244-1229]{Hubing Xiao}
\affiliation{Shanghai Key Lab for Astrophysics, Shanghai Normal University, Shanghai, 200234, China}

\author{Jianzhen Chen}
\affiliation{Shanghai Key Lab for Astrophysics, Shanghai Normal University, Shanghai, 200234, China}

\author{Anton A. Strigachev}
\affiliation{Institute of Astronomy and National Astronomical Observatory, Bulgarian Academy of Sciences, 72 Tsarigradsko shosse Blvd., 1784 Sofia, Bulgaria 
}

\author{Rumen S. Bachev}
\affiliation{Institute of Astronomy and National Astronomical Observatory, Bulgarian Academy of Sciences, 72 Tsarigradsko shosse Blvd., 1784 Sofia, Bulgaria 
}

\author{Xiangtao Zeng}
\affiliation{Center for Astrophysics, Guangzhou University, Guangzhou, 510006, China}
\affiliation{Key Laboratory for Astronomical Observation and Technology of Guangzhou, Guangzhou, 510006, China}
\affiliation{Astronomy Science and Technology Research Laboratory of Department of Education of Guangdong Province, Guangzhou, 510006, China}

\author{Marina Manganaro}
\affiliation{Department of Physics, University of Rijeka, Rijeka, 51000, Croatia}

\author{Rui Xue}
\affiliation{Department of Physics, Zhejiang Normal University, Jinhua, 321004, China}

\author{Zelin Li}
\affiliation{College of Microelectronics, Feicuihu Campus of Hefei University of Technology, Hefei, 230601, China}

\author[0000-0002-5929-0968]{Junhui Fan}
\affiliation{Center for Astrophysics, Guangzhou University, Guangzhou, 510006, China}
\affiliation{Key Laboratory for Astronomical Observation and Technology of Guangzhou, Guangzhou, 510006, China}
\affiliation{Astronomy Science and Technology Research Laboratory of Department of Education of Guangdong Province, Guangzhou, 510006, China}

\correspondingauthor{Junhui Fan, Jianzhen Chen, Hubing Xiao}
\email{fjh@gzhu.edu.cn, jzchen@shnu.edu.cn, hubing.xiao@shnu.edu.cn}



\begin{abstract}
The `blazar sequence' has been proposed for more than 20 years, yet its nature is still unclear.
In this work, for the first time, we expand this topic to the TeV band by using a sample of 58 TeV blazars including 48 blazars in the quiescent state and 21 blazars in the flaring state. 
We investigate the correlation between the TeV luminosity, which has been compensated for attenuation from extragalactic background light, and the synchrotron peak frequency. 
We note that there is no correlation between TeV luminosity and peak frequency in the quiescent state and a strong anti-correlation in the flaring state for the observed value. 
However, there is a strong positive correlation in both the quiescent state and the flaring state for the intrinsic value. 
This indicates that the blazar sequence is shown in the flaring state rather than in the quiescent state for the observed value and the blazar sequence is not present in both two states after removing the beaming effect. 
In addition, to confirm whether the beaming effect results in the blazar sequence, we compare the \textit{Fermi} $\gamma$-ray luminosity between the quiescent state and the flaring state. 
We find the \textit{Fermi} $\gamma$-ray luminosity in the flaring state is greater than that in the quiescent state and the Doppler factor in the flaring state is greater. 
We suggest the blazar sequence in the flaring state may be due to the stronger beaming effect.

\end{abstract}


% \keywords{galaxies: active --- BL Lacertae objects: general --- gamma-rays: galaxies}
\keywords{TeV blazars; BL Lacertae objects; Flat-spectrum radio quasars}


\section{Introduction} \label{sec:intro}

Blazars are a subclass of active galactic nuclei (AGNs) with extreme observational properties, including high polarization, fast rapid variability, and strong high-energy $\gamma$-ray radiation, etc (\citealp{Urry1995, Fan2004, Fan2014a, Fan2016, Lyutikov2017, Ouyang2021, Xiao2022}, and references therein). They are divided into two subclasses, flat spectrum radio quasars (FSRQs) and BL Lacertae objects (BL Lacs), based on their optical spectrum. 
FSRQs have strong broad emission lines (rest-frame equivalent width, EW $\ge$ 5 $\rm \mathring{A}$), while BL Lacs have weak or no emission lines (EW $<$ 5 $\rm \mathring{A}$) \citep{Stickel1991, Urry1995, Scarpa1997}.
The blazar spectral energy distribution (SED) is characterized by a two-hump structure. 
The low-energy hump occurs at infrared to X-ray explained by the synchrotron emission from electrons. 
The high-energy hump occurs at MeV to GeV which is produced either by the inverse Compton process in a leptonic scenario \citep{Tavecchio1998ApJ, Ghisellini2009MNRAS}, or the proton synchrotron emission in a hadronic scenario \citep{Mucke2003, Bottcher2009ApJ, Cerruti2017AIPC} even the hybrid scenario \citep{Cerruti2011, Bottcher2013}. 
According to the synchrotron peak frequency, blazars can be divided into low-synchrotron-peaked sources (LSPs), intermediate-synchrotron-peaked sources (ISPs), and high-synchrotron-peaked sources (HSPs) \citep{Abdo2010, Fan2016, Yang2022a}.

An increasing number of extragalactic sources are detected in the TeV band, which is also known as the very-high-energy band (VHE; $E \gtrsim \rm 100 ~GeV$). 
The radiation in the VHE band is mainly detected by ground-based Cherenkov telescopes, e.g., the Major Atmospheric Gamma-ray Imaging Cherenkov Telescopes (MAGIC), the High Energy Stereoscopic System (H.E.S.S.), the Collaboration between Australia and Nippon for a Gamma Ray Observatory in the Outback (CANGAROO), the Very Energetic Radiation Imaging Telescope Array System (VERITAS), and the Large High Altitude Air Shower Observatory (LHAASO). 
There are 251 TeV sources collected in TeVCat{\footnote{\url{http://tevcat2.uchicago.edu/}}} and the number is gradually increasing.
The number of TeV blazars is still small compared to GeV ones, but a lot of issues have been studied about the TeV blazar.
\citet{Massaro2013} suggested criteria to select TeV BL Lac candidates based on infrared and X-ray observations; 
\citet{Lin2016} compared the observational properties of TeV-detected BL Lacs with non-TeV-detected BL Lacs; 
\citet{Paiano2017} compiled 22 BL Lacs detected in TeV energy and determined or constrained their redshifts, etc. 
The observations of TeV blazars are usually performed during their flaring state, while some are observed in their low energy state (quiescent state) owing to certain circumstances, e.g., the source is very bright enough even in a quiescent state, etc.
However, observations of TeV blazar in VHE $\gamma$-rays are not easy because of the strong absorption by the extragalactic background light (EBL; \citealp{Franceschini2008, Dominguez2011}), the absorption would steepen the TeV spectrum significantly.

\citet{Fossati1998} proposed a blazar sequence, from FSRQs to radio-selected BL Lacs (RBLs) and to X-ray-selected BL Lacs (XBLs), whose synchrotron peak luminosity decreases with synchrotron peak frequency increasing, which has been explained as a consequence of the difference of electron cooling efficiency \citep{Ghisellini1998}.
The original blazar sequence was built with a few sources and \cite{Ghisellini2017} used the 3LAC (The 3rd Catalog of AGN Detected by the \textit{Fermi}-LAT) sample and compiled all the archival data to build their average SED, which agrees with the original blazar sequence and the existence of the blazar sequence is proven.
Many authors have performed many studies on this issue, and there are basically two elucidations among these scholars. 
One is that they acknowledged the validity of the blazar sequence, but not the same as the original explanation. 
For instance, it is found to be caused by a beaming effect \citep{Nieppola2008, Fan2017, Yang2022b} and the different location of the dissipation region for BL Lacs and FSRQs \citep{Potter2015}. 
The other one is that they argued that the blazar sequence is caused by observational biases \citep{Giommi2012a, Giommi2012b}.

The study of blazar sequence is still open, many of the previous works were carried out using monochromatic luminosity (radio, optical, X-ray, $\gamma$-ray, synchrotron peak luminosity) to study blazar sequence.
However, the blazar study in TeV band have not been performed yet.
Using the TeV $\gamma$-ray data to investigate the blazar sequence is of great significance to understand the physical interpretation of the blazar sequence in the very-high-energy band. 
In this work, for the purpose of exploring the blazar sequence in TeV energy range, we will investigate the correlation between the TeV luminosity and the synchrotron peak frequency for TeV blazars. 

The sample acquisition and quantity calculation are presented in Section \ref{Sample}. 
The results are given in Section \ref{Results} and Section \ref{fermi}. 
The discussions and conclusions are given in Section \ref{Discussions} and Section \ref{Conclusions}.  
In this work, we use the following flat $\Lambda$CDM cosmology, with $H_0 \rm = 71 ~km \cdot s^{-1}\cdot Mpc^{-1}$, $\Omega_{\rm M} = 0.27 $ and $\Omega_{\Lambda} = 0.73$.

\section{Sample}\label{Sample}
There are 81 blazars, including 67 BL Lacs, 9 FSRQs, and 5 blazar candidates of uncertain type (BCUs), confirmed as TeV emitters by associating TeVCat and 4FGL\_DR3 \citep{Abdollahi2022}. 
We collect the TeV spectrum information in literature and synchrotron peak frequency (${\rm log} \, \nu_{p}^{\rm ob}$) from \citet{Fan2016} and \citet{Yang2022a} for these sources. 
In this case, we compile a sample of 57 out of 81 TeV blazars with available the quiescent state or (and) flaring state TeV spectra. 
The quiescent state refer to a low flux density or being quiet and not active or being an average flux density state, while flaring state designates conspicuous transient brightening and shows an outburst in TeV band. 
Moreover, we also collect one source, 4FGL J2042.1+2427 (RGB J2042.1+2426; \citealp{2020ApJS..247...16A}), to our sample which was observed by MAGIC, but not listed in TeVCaT. 
Finally, we get a sample of 58 sources (49 BL Lacs, 7 FSRQs and 2 BCUs), of which 48 sources (44 BL Lacs, 2 FSRQs and 2 BCUs) are in a quiescent state and 21 sources (14 BL Lacs and 7 FSRQs) are in a flaring state.
These sources are listed in Table \ref{tab:quiescent} and Table \ref{tab:flaring} respectively. 

\subsection{The EBL correction}

The VHE spectrum of our sample is reduced by the EBL attenuation, making the de-absorbed spectrum information can not obtain directly. 
For this reason, we try to correct the absorbed spectrum.

Assuming the absorbed VHE spectrum of TeV blazar to form a power-law function:
\begin{equation}
\frac{\mathrm{d}N}{\mathrm{d}E}_{\rm abs} = N_{0, \rm abs} \times \left( \frac{E}{E_{0}}\right) ^{-\Gamma_{\rm abs}},
\end{equation}
where $\Gamma_{\rm abs}$ is a photon spectral index in VHE spectrum, $E_0$ is a normalization energy and $N_{0, \rm abs}$ is differential photon flux at $E_0$ in units of $\rm cm^{-2} \cdot s^{-1} \cdot TeV^{-1}$. 
Then the de-absorbed spectrum can be obtained by the following function:
\begin{equation}
\frac{\mathrm{d}N}{\mathrm{d}E}_{\rm de - abs} = \frac{\mathrm{d}N}{\mathrm{d}E}_{\rm abs} e^{\tau(E,z)} = N_{0, \rm de - abs} \times \left( \frac{E}{E_{0}}\right) ^{-\Gamma_{\rm de -abs}},
\end{equation}
where $\tau(E,z)$, the optical depth, is a function of redshift and energy. $N_{0, \rm de - abs}$ (hereafter $N_{0}$) is differential photon flux at $E_0$ and $\Gamma_{\rm de - abs}$ (hereafter $\Gamma$) is a photon spectral index with EBL de-absorption.
The model from \citet{Franceschini2008} is used to correct the VHE spectrum and then the de-absorbed spectrum information can be obtained. 
We consider the flux in TeV band in units of $\rm TeV \cdot cm^{-2} \cdot s^{-1}$, which can be calculated by
\begin{equation}
F = E^{2} \times \frac{\mathrm{d}N}{\mathrm{d}E}_{\rm de - abs} ,
\end{equation}
here $E$ is 1 TeV. 
Meanwhile, we show the absorbed (original) spectrum and the de-absorbed spectrum in  Figure \ref{fig:VHE_q} and Figure \ref{fig:VHE_f} in Appendix for the quiescent state and the flaring state. 

The isotropic TeV $\gamma$-ray luminosity in units of $\rm erg \cdot s^{-1}$ is expressed as
\begin{equation}
L=4\pi d_L^2 (1+z)^{\Gamma-2} F   {\rm ,} 
\label{cal_luminosity}
\end{equation}
here $(1+z)^{\Gamma-2}$ stands for a K-correction, $z$ is redshift, $\Gamma$ is the photon spectral index with EBL de-absorption, $F$ is the TeV $\gamma$-ray flux and $d_L$ represents a luminosity distance expressed by 
$d_{L}=\frac{(1+z)c}{H_0} \int_z^{1+z} \frac{1}{\sqrt{\Omega_{\rm M} x^3+1-\Omega_{\rm M}}} \mathrm{d}x $.
We calculate the de-absorbed TeV $\gamma$-ray luminosity ($L_{\rm 1 \, TeV , \, de-abs}^{\rm ob}$, hereafter $L_{\rm 1 \, TeV}^{\rm ob}$ ) at 1 TeV for the sources in our sample.

\section{Results}\label{Results}

The blazar sequence is reflected in the diagrams as an anti-correlation between luminosity in different bands and synchrotron peak frequency \citep{Fossati1998, Nieppola2006, Nieppola2008, Giommi2012a, Giommi2012b, Mao2016, Fan2017, Yang2022b}. 
Therefore, we can test the blazar sequence in the TeV band by studying the correlation between the TeV luminosity and synchrotron peak frequency.

In order to evaluate the correlation, the Pearson correlation analysis was used in this work.
This analysis presents the results in two parameters, namely correlation coefficient ($r$) and chance probability ($p$). 
The coefficient $r$, the most common way of measuring a linear correlation, is a number between –1 and 1 that measures the strength and direction of the relationship between two variables.
If the $p$ value is smaller than the critical value, which is usually set as 0.05, then the relationship is statistically significant and allows you to reject the null hypothesis, in which the two variables are not correlated.
In addition, the least squares linear regression was applied to find the linear correlation.

\subsection{The observed correlation between TeV luminosity and synchrotron peak frequency}

\begin{figure}[htbp]
    \centering
    \includegraphics[width=3.3 in]{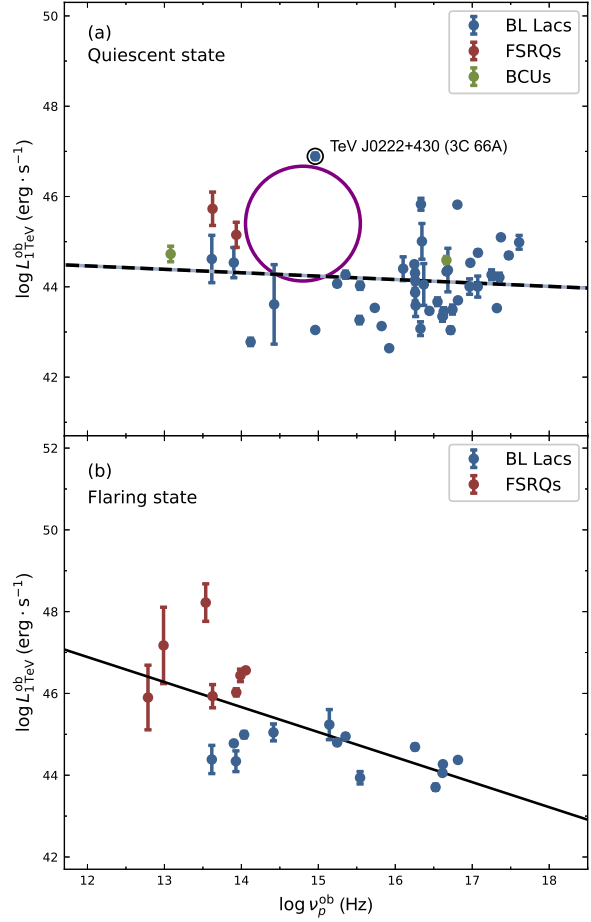}
    \caption{The observed correlation between ${\rm log} \, L^{\rm ob}_{\rm 1 \, TeV}$ and ${\rm log} \,\nu_{p}^{\rm ob}$ in both quiescent state and flaring state. (a) for the quiescent state, (b) for the flaring state.
    The blue circle is for BL Lacs, the red one is for FSRQs, and the green one is for BCUs. 
    The solid line represents the corresponding linear regression for the whole sample. 
    The violet circle represents the particular region to be discussed in Section \ref{Discussions} and the dashed line indicates a trend of the correlation.
    }
    \label{fig:observed}
\end{figure}

\textit{Quiescent state:} 
Based on the obtained TeV luminosity and synchrotron peak frequency, we plot ${\rm log} \, L^{\rm ob}_{\rm 1 \, TeV}$ vs ${\rm log} \, \nu^{\rm ob}_{p}$ in the top panel of Figure \ref{fig:observed}. 
When the Pearson correlation analysis is applied to test the correlation, we find that there is no significant correlation between ${\rm log} \, L^{\rm ob}_{\rm 1 \, TeV}$ and ${\rm log} \, \nu^{\rm ob}_{p}$ due to $r=-0.1$ and of $p= 50.39 \%$ for the case of TeV blazars in the quiescent state. 
However, we can still find that there is a trend of anti-correlation between the two parameters in this panel and reveal the existence of a blazar sequence, because some ISP TeV sources are absent in the circled region, which will be discussed in Section \ref{Discussions}.

\textit{Flaring state:} 
Compared to the quiescent state, we find an anti-correlation between ${\rm log} \, L^{\rm ob}_{\rm 1 \, TeV}$ and ${\rm log} \, \nu^{\rm ob}_{p}$ in the flaring state and present a linear regression:
\begin{equation}
{\rm log} \, L^{\rm ob}_{\rm 1 \, TeV} = -(0.61 \pm 0.15) ~{\rm log} \, \nu^{\rm ob}_{p} + (54.23 \pm 2.27) 
\end{equation}
with the $r=-0.67$ and $p=7.95 \times 10^{-4}$, as shown in the lower panel of Figure \ref{fig:observed}. 
This result suggests a strong anti-correlation between TeV luminosity and the synchrotron peak frequency in the flaring state and it indicates that the blazar sequence is present in the flaring state rather than in the quiescent state for TeV blazars.

\subsection{The intrinsic correlation between TeV luminosity and synchrotron peak frequency}
In a relativistic beaming model, the observed flux density $f^{\rm ob}$ and the intrinsic one $f^{\rm in}$ exist such correlation: $f^{\rm ob}=\delta^{q} f^{\rm in}$ ($\delta = \left[ \gamma_{\rm bulk}(1-\beta_{\rm app} \cos \theta) \right] ^{-1}$, where $\delta$ is a Doppler factor, $\beta_{\rm app}$ is the apparent velocity, $\theta$ is the viewing angle and $\gamma_{\rm bulk}$ is bulk Lorentz factor), 
then the intrinsic luminosity ($L^{\rm in}$) and frequency ($\nu^{\rm in}$) can be calculated though the observed luminosity ($L^{\rm ob}$) and frequency ($\nu^{\rm ob}$): 
\begin{equation} 
L^{\rm in}=\frac{L^{\rm ob}}{\delta^{q+1}},~ \nu^{\rm in}=\frac{\nu^{\rm ob}}{\delta/(1+z)}  {\rm ,}
\end{equation}
where $q = 2 + \alpha $ for a continuous jet, $q = 3 + \alpha$ for a spherical jet and $\alpha$ is the spectral index ($f_{\nu} \propto \nu^{-\alpha}, \, \alpha = \Gamma-1$) \citep{Lind1985}. 
We compile the Doppler factors from \cite{Chen2018} and the values range from 1.3 to 67.5 for the quiescent state and from 1.6 to 67.5 for the flaring state in our sample. 
We calculate the intrinsic TeV luminosity $L_{\rm 1 \, TeV}^{\rm in}$ at 1 TeV for the sources in our sample with available Doppler factor for both two states. 
The remaining sources with unavailable Doppler factors are excluded in the analysis of the intrinsic correlation.

\begin{figure}[htbp]
    \centering
    \includegraphics[width=5.8 in]{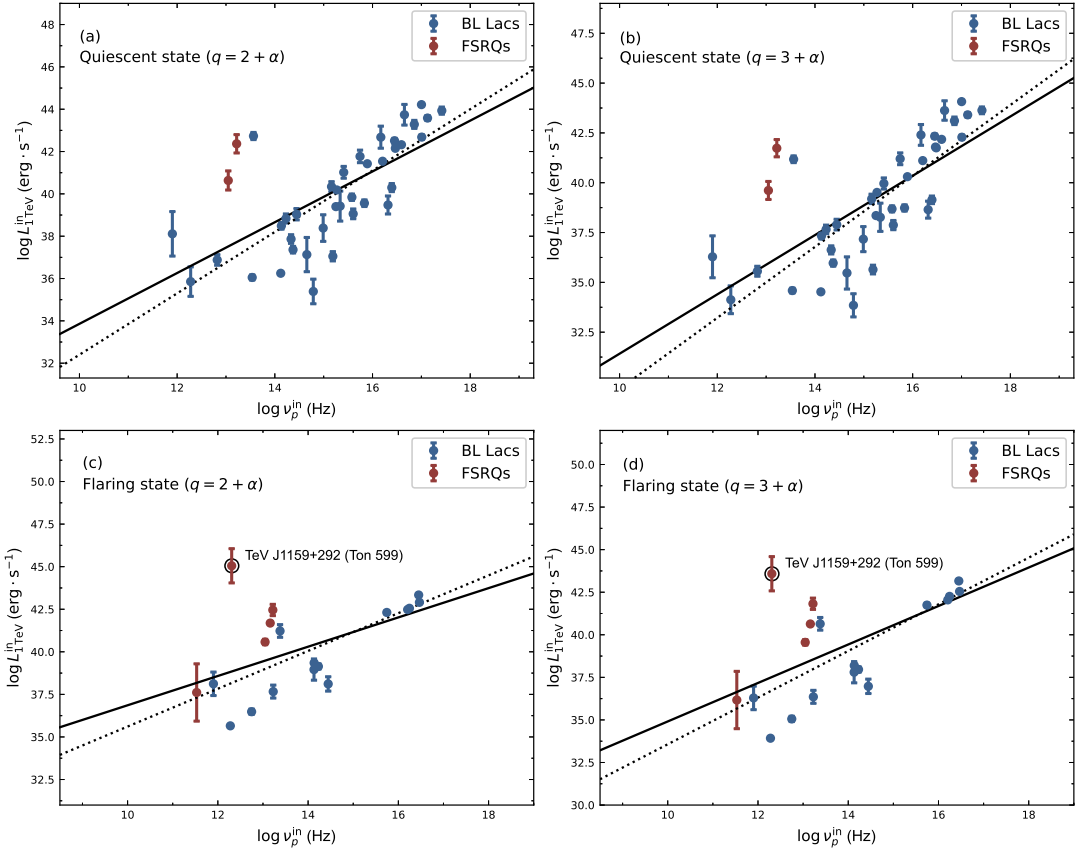}
    \caption{The intrinsic correlation between ${\rm log} \, L^{\rm in}_{\rm 1 \, TeV}$ and ${\rm log}\, \nu_{p}^{\rm in}$ in both quiescent state and flaring state.
    (a) and (b) for the quiescent state, (c) and (d) for the flaring state; (a) and (c) for the case of $q = 2 + \alpha$, (b) and (d) for the case of $q = 3 + \alpha$. 
    The blue circle is for BL Lacs and the red one is for FSRQs. 
    The solid lines represent the corresponding linear regression for the whole sample. 
    The dotted lines will be discussed in Section \ref{Discussions}. 
    }
    \label{fig:intrinsic}
\end{figure}

\textit{Quiescent state:} 
The top panel of Figure \ref{fig:intrinsic} shows a positive correlation between the intrinsic TeV luminosity and the synchrotron peak frequency in the quiescent state. We obtain
\begin{equation}
{\rm log} \, L^{\rm in}_{\rm 1 \, TeV} = (1.20 \pm 0.21) ~{\rm log} \, \nu^{\rm in}_{p} + (21.86 \pm 3.23) ,
\end{equation}
with $r=0.67$ and $p \, \textless \, 10^{-4}$ for the case of $q = \alpha +2$. 
And 
\begin{equation}
{\rm log} \, L^{\rm in}_{\rm 1 \, TeV} = (1.49 \pm 0.24) ~{\rm log} \, \nu^{\rm in}_{p} + (16.54 \pm 3.66) ,
\end{equation}
with $r=0.71$ and $p \, \textless \, 10^{-4}$ for the case of $q = \alpha +3$.

\textit{Flaring state:} 
The lower panel of Figure \ref{fig:intrinsic} also shows a positive correlation between the intrinsic TeV luminosity and the synchrotron peak frequency in the flaring state. 
We obtain 
\begin{equation}
{\rm log} \, L^{\rm in}_{\rm 1 \, TeV} = (0.86 \pm 0.34) ~{\rm log} \, \nu^{\rm in}_{p} + (28.27 \pm 4.72) ,
\end{equation}
with $r=0.53$ and $p=2.02 \%$ for the case of $q = \alpha +2$. 
And 
\begin{equation}
{\rm log} \, L^{\rm in}_{\rm 1 \, TeV} = (1.13 \pm 0.36) ~{\rm log} \, \nu^{\rm in}_{p} + (23.61 \pm 5.05) ,
\end{equation}
with $r=0.60$ and $p = 6.17 \times 10^{-3}$ for the case of $q = \alpha +3$.  

From the intrinsic relationship, we note that the intrinsic TeV luminosity and synchrotron peak frequency show a very strong positive correlation in both quiescent and flaring states. 
Thus, we suggest that the blazar sequence is not present in the TeV band after removing the beaming effect. 
As above-mentioned, we notice the absence (or only a trend) of the blazar sequence for the quiescent state and the presence of a significant blazar sequence for the flaring state in Figure \ref{fig:observed}. 
Moreover, the flatter slopes of intrinsic diagrams for the flaring state than that for the quiescent state in Figure \ref{fig:intrinsic} indicate a stronger beaming effect for the flaring state. 
And we shall discuss this issue in Section \ref{fermi}. 

\section{The comparison of \textit{Fermi} $\gamma$-ray emission between the quiescent state and flaring state} \label{fermi}
In order to investigate whether the stronger beaming effect results in (or boosts) the presence of the blazar sequence in the flaring state, we try to compare the $\gamma$-ray emission between the quiescent state and flaring state for the sources in our sample. 
Previous studies have found that the bulk Lorentz factor (or the Doppler factor, $\delta \approx \gamma_{\rm bulk}$ for blazar) shows a strong correlation with $\gamma$-ray emission for the \textit{Fermi} detected AGNs \citep{Fan2014b, Ghisellini2014, Blandford2019, Zhang2020}. 
Therefore, the $\gamma$-ray emission can be considered to be an indicator of the beaming effect. 

\subsection{Introduction to \textit{Fermi}-LAT and data analysis }

The \textit{Fermi}-LAT\footnote{\url{https://fermi.gsfc.nasa.gov}} detects $\gamma$-ray in the energy from 20 MeV to beyond 300 GeV and surveys the entire sky every day \citep{Atwood2009}. 
In order to obtain the GeV $\gamma$-ray emission in the quiescent state and flaring state, we analyse \textit{Fermi} data in the same observation periods when the VHE observation is performing. 
We use MJD 57863 as the starting time of the data for analysis if the VHE observation period encompasses the campaign before the launch of the \textit{Fermi}-LAT. 
For TeV J1443+120, for instance, the observation period spans the months of May 2008 and June 2010, totaling 53 hours and then we choose MJD 57863 as its start time. 
We perform the binned likelihood analysis of the data using the latest LAT \textsl{Fermitools}{\footnote{\url{https://fermi.gsfc.nasa.gov/ssc/data/analysis/}}} 2.2.0 and the instrument response functions \textsl{P8R3\_SOURCE\_V3}{\footnote{\url{https://fermi.gsfc.nasa.gov/ssc/data/analysis/documentation/Cicerone/Cicerone\_LAT\_IRFs/IRF\_overview.html}}}. 
The energy of the data ranges from 0.1 GeV to 300 GeV and we select the maximum zenith angle value as 90° to reduce the photons coming from the Earth limb. 
The condition `evclass = 128, evtype = 3' is used to filter events with high probability of being photons.
For each source, the region of interest (ROI) considered is the circle of 10° radius surrounding the catalog source position. 
The condition `(DATA\_QUAL $\ge$ 0)\&\&(LAT\_CONFIG==1)' is used to select the good time intervals. 
The background model consists of the Galactic (\textsl{gll\_iem\_v07.fits}) and isotropic extragalactic diffuse emission models (\textsl{iso\_P8R3\_SOURCE\_V3\_v1.txt}{\footnote{\url{https://fermi.gsfc.nasa.gov/ssc/data/access/lat/BackgroundModels.html}}}). 
The normalizations of the two diffuse emission components are set as free parameters in the analysis. When performing the likelihood fitting, we fix the parameters of the least significant sources until convergence is reached.

The spectrum of each source is described with a log-parabola function 
\begin{equation}
    \frac{{\rm d}N}{{\rm d}E} = N_{0} \times (\frac{E}{E_{0}})^{- \Gamma -\beta \log(\frac{E}{E_{0}})}  {\rm ,} 
\end{equation}
where $N_{0}$ is the normalization in units of $\rm cm^{-2} \cdot s^{-1} \cdot MeV^{-1}$, $E_{0}$ is the pivot energy and $\Gamma$ is photon index at $E_{0}$ and $\beta$ is the curvature index,
or a power-law function 
\begin{equation}
\frac{{\rm d}N}{{\rm d}E} = N_{0} \times (\frac{E}{E_{0}})^{-\Gamma}  {\rm ,}      
\end{equation}
where $N_{0}$ is the normalization in units of $\rm cm^{-2} \cdot s^{-1} \cdot MeV^{-1}$, $E_{0}$ is the pivot energy and $\Gamma$ is photon index.
After we obtain the normalisation and the parameters, the integral flux ($F$) in units of $\rm GeV \cdot cm^{-2} \cdot s^{-1}$ can be expressed in the form 
\begin{equation}
    F = \int_{E_{\rm L}}^{E_{\rm U}}  E \times \frac{{\rm d}N}{{\rm d}E} \, {\rm d}E
\end{equation}
for the log-parabola or the power-law function. 
Here $E_{\rm L}$ and $E_{\rm U}$ are corresponding to 1 GeV and 100 GeV, respectively. 
Then we calculate the $\gamma$-ray luminosity ($L_{\gamma}$) in units of $\rm erg \cdot s^{-1}$ through the Equation (\ref{cal_luminosity}).

\subsection{The results and conclusions obtained from the \textit{Fermi}-LAT data analysis}
\begin{figure}[htbp]
    \centering
    \includegraphics[width=4 in]{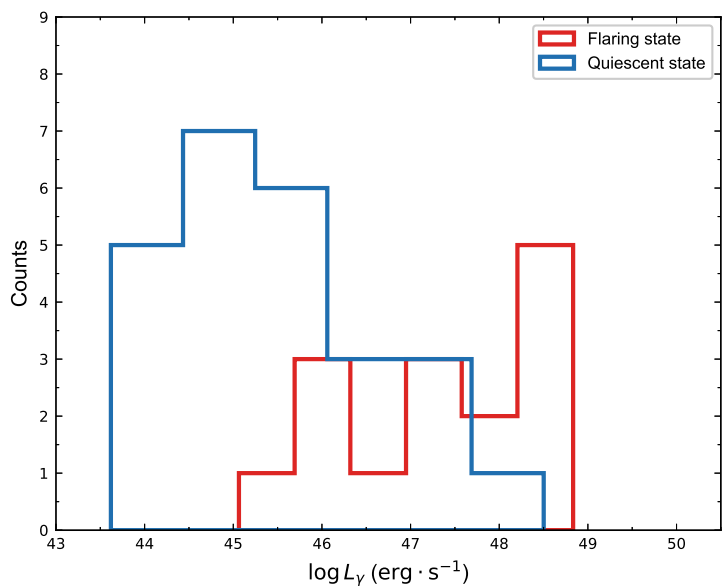}
    \caption{The distribution of \textit{Fermi} $\gamma$-ray luminosity in both quiescent state and flaring state. The blue histogram denotes the quiescent state and red histogram denotes the flaring state.}  
    \label{fig:L_fermi_dis}
\end{figure}

Finally, we obtain the sample of \textit{Fermi} $\gamma$-ray luminosity in the quiescent state and flaring state, with sample size 25 (22 BL Lacs, 2 FSRQs and 1 BCU) and 15 (9 BL Lacs and 6 FSRQs), respectively. 
We exclude the sources which were not significantly detected with a test statistic TS $<$ 9 and/or the model-predicted photons $N_{\rm pred} <$ 8 \citep{Kapanadze2022}. 
The fitting normalizations and parameters of each source are listed in Table \ref{tab:fermi_quiescent} and Table \ref{tab:fermi_flaring}. 
The \textit{Fermi} $\gamma$-ray luminosity distribution of both quiescent state and flaring state is shown in Figure \ref{fig:L_fermi_dis}. 
We compute the mean value of $\gamma$-ray luminosity of both two states and get the mean value of $\log L_{\gamma, \, \rm flaring}^{\rm mean} = 47.37 \pm 0.30 {\, \rm erg \cdot s^{-1}}$ and $ \log L_{\gamma, \, \rm quiescent}^{\rm mean} = 45.51 \pm 0.24 {\, \rm erg \cdot s^{-1}}$, respectively. 
A Mann-Whitney U test is employed to compare the \textit{Fermi} $\gamma$-ray luminosity distribution of these two states, the test gives a statistical value of 327.0 and a $p$-value smaller than $10^{-4}$, which suggests that their mean values are different and the \textit{Fermi} $\gamma$-ray luminosity of the flaring state is greater than that of the quiescent state. 
This result also reveals that blazars have a stronger beaming effect during the flaring state than that during the quiescent state, therefore, we suggest the blazar sequence may be caused by the stronger beaming effect for those sources observed in the flaring state. 

%%%%%%%%%%%%%%%%%%%%%%%%%%%%%%%%%%%%%%%%%%%%%%%%%%%%%%%%%%%%%%%%%%%%%%%%%%
\section{Discussions}\label{Discussions}

\citet{Fossati1998} calculated the SEDs for 126 blazars, investigated the correlation between radio (5 GHz) luminosity and synchrotron peak luminosity, between $\gamma$-ray luminosity and synchrotron peak frequency, and proposed the `blazar sequence'.
This sequence was later confirmed with a larger sample of \textit{Fermi} sources \citep{Ghisellini2017}.

The synchrotron radiation of blazars is produced by the relativistic electrons moving in the magnetic field. 
The $\gamma$-ray emission is attributed to the inverse Compton process, but different seed photons supply for FSRQs and for BL Lacs.
FSRQs are considered to contain a soft-photon-rich environment, these soft photons could come from the accretion disk, the broad emission line region (BLR), the dusty torus (DT) \citep{Ghisellini1998, Prandini2022}.
Hence, the electron cooling for FSRQs is more efficient than that for BL Lacs and the energy of relativistic electrons in BL Lac jets is more likely powerful than those in FSRQ jets. 
As a result, the energy of soft photons produced through synchrotron radiation in BL Lac jets is higher than that in FSRQ jets, making BL Lac have a higher synchrotron peak frequency \citep{Ghisellini1998}.
These seed photons could be scattered into an extremely high energy, e.g., TeV range, mainly by the population of relativistic electrons through the synchrotron-self-Compton (SSC) process for BL Lacs, and through the external Compton (EC) process for FSRQs. 
The more dramatic electron cooling efficiency makes the lower energy of relativistic electrons and the subsequent lower synchrotron peak frequency. 
However, BL Lacs have a less efficient electron cooling process, the relativistic electron could preserve energy and result in higher energy of synchrotron peak and produce SSC TeV emission.

In this work, we expand the study of the `blazar sequence' to the TeV band.
We notice that there is an anti-correlation between the observed TeV luminosity and the observed synchrotron peaked frequency during the flaring state,
and there is no significant anti-correlation between the observed TeV luminosity and the observed synchrotron peaked frequency for the sources in the quiescent state, as seen in Figure \ref{fig:observed}.
In the top panel of Figure \ref{fig:observed}, we consider and analyse the correlation for FSRQs, BL Lacs and BCUs together as a whole sample, because we have an extremely limited sample size of FSRQs and BCUs.
Besides, the empty circled region, synchrotron peak frequency ranging from $10^{14}$ up to $10^{16}$, in this panel is intriguing.
Those sources located in this region should be mostly considered as ISPs with intermediate TeV luminosity compare to TeV J0222+430 (3C 66A) and those sources below the circle.
These sources, somehow, in this region are not detected.
Therefore, the anti-correlation is diluted in the quiescent state.
In this case, our results indicate a trend of blazar sequence at the TeV band in the quiescent state of blazars and the trend is plotted in a dashed line in the top panel of Figure \ref{fig:observed}. 
However, one should bear in mind that it is possible that there are no more other sources in this empty circle.
Because these potential sources are detectable to those TeV telescopes.
In this case, the anti-correlation trend for the quiescent state does not exist, and the TeV blazar sequence appears only in the flaring state. 
To sum up, the existence of the TeV blazar sequence in the quiescent state is uncertain.

\citet{Fan2017} studied the blazar sequence in radio, optical, X-ray and $\gamma$-ray bands using a sample of 86 flaring blazars.
They noticed that the blazar sequence exists in the diagram of observed quantities and the sequence disappears when the intrinsic quantities are considered, suggesting that the blazar sequence is a consequence of a beaming effect. 
\citet{Yang2022b}, an extended work of \citet{Fan2017} with a larger sample, also found the blazar sequence exists for the observed quantities and disappears for the intrinsic quantities. 
Our results in Figure \ref{fig:intrinsic} are consistent with the conclusion of \citet{Fan2017} and \citet{Yang2022b}. 
In the top panel of Figure \ref{fig:intrinsic}, we notice that there are two FSRQs mixed with those BL Lacs.
We reanalysed the correlation after removing the two FSRQs, the linear regressions are plotted in the top panel of Figure \ref{fig:intrinsic} with dotted lines and give the formulas
\begin{equation}
    {\rm log} \, L^{\rm in}_{\rm 1 \, TeV} = (1.45 \pm 0.20) ~{\rm log} \, \nu^{\rm in}_{p} + (17.90 \pm 3.03)
\end{equation}
with $r=0.77$ and $p \, \textless \, 10^{-4}$ for the case of $q=2+\alpha$; 
\begin{equation}
    {\rm log} \, L^{\rm in}_{\rm 1 \, TeV} = (1.78 \pm 0.22) ~{\rm log} \, \nu^{\rm in}_{p} + (11.86 \pm 3.38)
\end{equation}
with $r=0.80$ and $p \, \textless \, 10^{-4}$ for the case of $q=3+\alpha$. 
For the lower panel of Figure \ref{fig:intrinsic}, we notice TeV 1159+292 (Ton 599), shows an extremely powerful TeV emission, apart from the flaring sample population, where linearly distributed along the regression line.
Ton 599 was occurring in an extraordinary $\gamma$-ray flaring state.
This flare was detected by AGILE (the Astro-rivelatore Gamma a Immagini Leggero) from 16 to 18 December 2017 (MJD 58103.24 - 58105.24) \citep{Bulgarelli2017} and VERITAS from 15 to 16 December 2017 (MJD 58102 - 58104) \citep{Mukherjee2017}.
The average flux on the two nights is $F(E \ge 100 \, \rm GeV)=(1.0 \pm 0.1) \times 10^{-10} \, \rm ph \cdot cm^{-2} \cdot s^{-1}$ or 16\% of the Crab Nebula flux above the same threshold \citep{Mukherjee2017}. 
According to our fitting results for the VHE spectrum, we find the photon spectral index of Ton 599 is $\Gamma = 0.16$ (see Figure \ref{fig:VHE_f}) and the spectrum is so hard that it has an extremely high luminosity in TeV emission. 

Thus, we consider reanalysing the linear regression without Ton 599, the results are plotted in the lower panel of Figure \ref{fig:intrinsic} in dotted lines and expressed as 
\begin{equation}
    {\rm log} \, L^{\rm in}_{\rm 1 \, TeV} = (1.11 \pm 0.26) ~{\rm log} \, \nu^{\rm in}_{p} + (24.51 \pm 3.65)
\end{equation}
with $r=0.73$ and $p = 5.80 \times 10^{-4}$ for the case of $q=2+\alpha$;
\begin{equation}
    {\rm log} \, L^{\rm in}_{\rm 1 \, TeV} = (1.37 \pm 0.29) ~{\rm log} \, \nu^{\rm in}_{p} + (19.87 \pm 4.15)
\end{equation}
with $r=0.76$ and $p = 2.65 \times 10^{-4}$ for the case of $q=3+\alpha$. 

It is clear that the correlations become more significant when we removed these particular sources.
And, our results of Figure \ref{fig:intrinsic} indicate the TeV blazar sequence is caused by the beaming effect which is consistent with the results in \citet{Fan2017} and \citet{Yang2022b}.
Moreover, we find these TeV sources have larger average \textit{Fermi} GeV $\gamma$-ray luminosity during the TeV flaring state than during the quiescent state, see in Figure \ref{fig:L_fermi_dis}.
These results suggest blazars in flaring state show a stronger beaming effect and show a more significant blazar sequence than those blazars in quiescent state.

\section{Conclusions}\label{Conclusions}
In this work, for the purpose of studying the blazar sequence in the TeV energy range, 
We have collected a sample of 48 TeV blazars in the quiescent state and 21 TeV blazars in the flaring state with available VHE spectrum information, synchrotron peak frequency, and Doppler factor from the literature. 
We have investigated the correlation, which is representative of the blazar sequence, between the observed/intrinsic TeV luminosity and the observed/intrinsic synchrotron peak frequency. 

Our main conclusions are as follows: 
\begin{itemize}
  \item [(1)] 
  There is no correlation between TeV luminosity and synchrotron peak frequency for observed value in the quiescent state, however, there is a strong anti-correlation between these two parameters in the flaring state. 
  Therefore, we suggest that the TeV blazar sequence is present in the flaring state.  
  \item [(2)]
  The correlation for intrinsic value shows a strong positive correlation in both the quiescent state and flaring state. 
  This result demonstrates that the blazar sequence is not present for the intrinsic value and the anti-correlation for the observed quantities derives from the beaming effect. 
  \item [(3)]
  Through the comparison of \textit{Fermi} $\gamma$-ray luminosity between the quiescent state and flaring state, we find that the \textit{Fermi} $\gamma$-ray luminosity of the flaring state is greater than that of the quiescent state, which means the Doppler factor in the flaring state is greater than that in the quiescent state. 
  We suggest the presence of the blazar sequence or the anti-correlation between TeV luminosity and synchrotron peak frequency may be due to the stronger beaming effect in the flaring state. 
\end{itemize}

\begin{acknowledgments}
We thank the support of the key laboratory for astrophysics of Shanghai.
H. B, Xiao acknowledges the support from the National Natural Science Foundation of China (NSFC 12203034) and from the Shanghai Science and Technology Fund (22YF1431500);
J. H, Fan acknowledges the support from the NSFC (NSFC U2031201, NSFC 11733001, U2031112), Scientific and Technological Cooperation Projects (2020-2023) between the People's Republic of China and the Republic of Bulgaria, Guangdong Major Project of Basic and Applied Basic Research (Grant No. 2019B030302001), the science research grants from the China Manned Space Project with NO. CMS-CSST-2021-A06, and the support for Astrophysics Key Subjects of Guangdong Province and Guangzhou City. 
This research was partially supported by the Bulgarian National Science Fund of the Ministry of Education and Science under grants KP-06-H28/3 (2018), KP-06-H38/4 (2019), KP-06-KITAJ/2 (2020) and KP-06-PN68/1(2022). 

\end{acknowledgments}

\bibliography{Ref}{}
\bibliographystyle{aasjournal}

\begin{deluxetable*}{lcccccccccccccccc}
\tabletypesize{\tiny}
\tablewidth{0pt} 
% %\tablenum{1}
\tablecaption{The sample of 48 TeV blazars in the quiescent state.\\ \label{tab:quiescent}}
\rotate 
\tablehead{ 
 TeV name & Other name & $z$ & Class & $\log \, \nu_{p}^{\rm ob}$ & $\delta$ & $N_{0, \rm abs}$   & $\Delta N_{0, \rm abs}$ & $\Gamma_{\rm abs}$ & $\Delta \Gamma_{\rm abs}$ & $E_{0}$ & $N_{0}$ & $\Delta N_{0}$ & $\Gamma$ & $\Delta \Gamma$ & $E_{0}$ & Ref.  \\
}
\colnumbers
\startdata
   TeV J0013-188 & SHBL J001355.9-185406 & 0.094 & B   & 14.96 & 29    & 1.20E-12 & 2.09E-14 & 3.43  & 0.05  & 0.51  & 1.94E-12 & 5.98E-14 & 2.78  & 0.08  & 0.51  & \cite{2013AA...554A..72H} \\
    TeV J0033-193 & KUV 00311-1938 & 0.61  & B   & 16.81 & 13.2  & 3.14E-11 & 3.00E-27 & 4.39  & 2.60E-16  & 0.17  & 8.37E-11 & 1.47E-26 & 1.92  & 4.77E-16  & 0.17  & \cite{2020MNRAS.494.5590A} \\
    TeV J0112+227 & S2 0109+22 & 0.265 & B   & 13.90 & 53.4  & 1.56E-10 & 1.86E-11 & 3.39  & 0.22  & 0.16  & 4.84E-10 & 7.47E-11 & 3.03  & 0.35  & 0.12  & \cite{2018MNRAS.480..879M} \\
    TeV J0152+017 & RGB J0152+017 & 0.08  & B   & 15.73 & 24.8  & 6.24E-13 & 5.81E-14 & 2.95  & 0.12  & 1     & 1.34E-12 & 1.14E-13 & 2.43  & 0.11  & 1     & \cite{2008AA...481L.103A} \\
    TeV J0214+517 & TXS 0210+515 & 0.049 & B   & 15.92 & 4.7   & 1.32E-13 & 3.28E-15 & 1.95  & 0.03  & 1.574 & 2.38E-13 & 1.47E-15 & 1.65  & 0.01  & 1.574 & \cite{2020ApJS..247...16A} \\
    TeV J0222+430 & 3C 66A & 0.444 & B   & 14.96 & 36    & 1.67E-11 & 2.24E-12 & 3.08  & 0.18  & 0.3   & 2.48E-10 & 2.49E-11 & 0.67  & 0.10  & 0.3   & \cite{2009ApJ...692L..29A} \\
    TeV J0232+202 & 1ES 0229+200 & 0.139 & B   & 16.26 & 14.2  & 6.74E-13 & 9.44E-14 & 2.51  & 0.15  & 1     & 2.70E-12 & 4.08E-13 & 1.44  & 0.20  & 1     & \cite{2007AA...475L...9A} \\
    TeV J0303-241 & PKS 0301-243 & 0.2657 & B   & 16.10 & 16.4  & 1.09E-11 & 1.85E-12 & 4.48  & 0.62  & 0.29  & 2.27E-11 & 2.83E-12 & 2.95  & 0.47  & 0.29  & \cite{2013AA...559A.136H} \\
    TeV J0319+187 & RBS 0413 & 0.19  & B   & 16.98 & 3.7   & 1.37E-11 & 1.11E-12 & 3.18  & 0.15  & 0.3   & 2.10E-11 & 8.82E-13 & 1.89  & 0.08  & 0.3   & \cite{2012ApJ...750...94A} \\
    TeV J0349-119 & 1ES 0347-121 & 0.188 & B   & 17.47 & 14.3  & 4.52E-13 & 6.97E-14 & 3.10  & 0.18  & 1     & 3.27E-12 & 4.29E-13 & 1.81  & 0.16  & 1     & \cite{2007AA...473L..25A} \\
    TeV J0416+010 & 1ES 0414+009 & 0.287 & B   & 17.61 & 2     & 5.54E-12 & 6.33E-13 & 3.29  & 0.15  & 0.305 & 1.59E-11 & 2.91E-12 & 1.50  & 0.25  & 0.305 & \cite{2012AA...538A.103H} \\
    TeV J0449-438 & PKS 0447-439 & 0.205 & U   & 13.08 &   --    & 2.81E-13 & 1.13E-13 & 3.99  & 0.44  & 1     & 2.53E-12 & 9.81E-13 & 2.57  & 0.43  & 1     & \cite{2013AA...552A.118H} \\
    TeV J0509+056 & TXS 0506+056 & 0.3365 & B   & 14.43 &   --    & 6.42E-11 & 1.16E-11 & 4.81  & 0.91  & 0.15  & 8.20E-11 & 1.68E-11 & 4.05  & 1.05  & 0.15  & \cite{2018ApJ...861L..20A} \\
    TeV J0521+211 & VER J0521+211 & 0.108 & B   & 15.24 & 14.3  & 1.99E-11 & 1.27E-12 & 3.46  & 0.14  & 0.4   & 3.08E-11 & 1.70E-12 & 2.83  & 0.13  & 0.4   & \cite{2013ApJ...776...69A} \\
    TeV J0648+152 & RX J0648.7+1516 & 0.179 & B   & 17.07 & 6.7   & 2.32E-11 & 2.71E-12 & 4.37  & 0.43  & 0.3   & 3.74E-11 & 4.30E-12 & 3.48  & 0.43  & 0.3   & \cite{2011ApJ...742..127A} \\
    TeV J0710+591 & RGB J0710+591 & 0.125 & B   & 17.25 & 1.5   & 5.58E-12 & 1.03E-12 & 2.70  & 0.20  & 0.5   & 1.08E-11 & 2.16E-12 & 1.88  & 0.22  & 0.5   & \cite{2010ApJ...715L..49A} \\
    TeV J0809+523 & 1ES 0806+524 & 0.138 & B   & 16.26 & 3.7   & 3.90E-12 & 7.56E-13 & 2.77  & 0.25  & 0.5   & 7.77E-12 & 2.33E-12 & 2.18  & 0.41  & 0.5   & \cite{2015MNRAS.451..739A} \\
    TeV J0847+115 & RBS 723 & 0.198 & B   & 16.37$^{*}$ & 1.9   & 1.10E-11 & 2.00E-12 & 3.76  & 0.71  & 0.3   & 1.84E-11 & 3.96E-12 & 2.93  & 0.86  & 0.3   & \cite{2020ApJS..247...16A} \\
    TeV J0854+201 & OJ 287 & 0.3056 & B   & 13.62 & 67.5  & 3.20E-11 & 6.45E-12 & 2.24  & 0.29  & 0.1   & 3.53E-11 & 1.23E-11 & 1.56  & 0.50  & 0.1   & \cite{2009PASJ...61.1011S} \\
    TeV J1010-313 & 1RXS J101015.9-311909 & 0.142639 & B   & 16.26 & 45.8  & 2.04E-13 & 8.68E-14 & 3.00  & 0.46  & 1     & 8.71E-13 & 3.71E-13 & 2.06  & 0.47  & 1     & \cite{2012AA...542A..94H} \\
    TeV J1015+494 & 1ES 1011+496 & 0.212 & B   & 16.68 & 1.3   & 1.97E-10 & 3.47E-11 & 4.11  & 0.51  & 0.2   & 2.60E-10 & 5.50E-11 & 3.56  & 0.67  & 0.2   & \cite{2007ApJ...667L..21A} \\
    TeV J1103-234 & 1ES 1101-232 & 0.186 & B   & 17.07 & 1.4   & 5.52E-13 & 7.95E-14 & 2.94  & 0.19  & 1     & 3.92E-12 & 4.90E-13 & 1.68  & 0.16  & 1     & \cite{2007AA...470..475A} \\
    TeV J1104+382 & Mkn 421 & 0.031 & B   & 16.61 & 1.5   & 1.58E-10 & 1.48E-11 & 2.89  & 0.19  & 0.3   & 1.70E-10 & 1.58E-11 & 2.74  & 0.19  & 0.3   & \cite{2016ApJ...819..156B} \\
    TeV J1136+701 & Mkn 180 & 0.045 & B   & 16.72 & 1.4   & 4.27E-11 & 2.58E-12 & 3.16  & 0.08  & 0.3   & 4.75E-11 & 3.90E-12 & 2.91  & 0.11  & 0.3   & \cite{2006ApJ...648L.105A} \\
    TeV J1217+301 & 1ES 1215+303 & 0.131 & B   & 15.35 & 15.1  & 1.53E-11 & 1.28E-12 & 3.29  & 0.16  & 0.4   & 2.66E-11 & 2.27E-12 & 2.61  & 0.17  & 0.4   & \cite{2020ApJ...891..170V} \\
    TeV J1221+282 & W Comae & 0.103 & B   & 15.54 & 13.8  & 7.35E-12 & 7.10E-13 & 3.06  & 0.21  & 0.5   & 1.26E-11 & 1.23E-12 & 2.38  & 0.22  & 0.5   & \cite{2009ApJ...695.1370A} \\
    TeV J1221+301 & 1ES 1218+304 & 0.182 & B   & 17.37 &   --    & 1.15E-11 & 7.81E-13 & 3.07  & 0.10  & 0.5   & 3.39E-11 & 1.76E-12 & 1.98  & 0.08  & 0.5   & \cite{2010ApJ...709L.163A} \\
    TeV J1315-426 & 1ES 1312-423 & 0.105 & B   & 16.44 &   --    & 2.40E-13 & 2.51E-14 & 2.68  & 0.15  & 1     & 6.68E-13 & 7.73E-14 & 2.00  & 0.17  & 1     & \cite{2013MNRAS.434.1889H} \\
    TeV J1422+323 & B2 1420+32 & 0.682 & F  & 13.63 & 4.3   & 1.04E-09 & 4.23E-11 & 4.19  & 0.09  & 0.1   & 1.70E-09 & 2.52E-10 & 3.31  & 0.36  & 0.1   & \cite{2021AA...647A.163M} \\
    TeV J1427+238 & PKS 1424+240 & 0.16  & B   & 16.26 & 34.8  & 5.01E-11 & 4.67E-12 & 3.85  & 0.27  & 0.2   & 6.05E-11 & 6.62E-12 & 3.32  & 0.34  & 0.2   & \cite{2010ApJ...708L.100A} \\
    TeV J1428+426 & 1ES 1426+428 & 0.129 & B   & 17.35 & 2.5   & 2.47E-11 & 3.69E-12 & 2.50  & 0.16  & 0.242 & 3.29E-11 & 3.62E-12 & 1.84  & 0.13  & 0.242 & \cite{2020ApJS..247...16A} \\
    TeV J1443+120 & 1ES 1440+122 & 0.16306 & B   & 16.97 & 1.5   & 1.50E-12 & 3.24E-13 & 3.09  & 0.44  & 0.5   & 3.75E-12 & 8.12E-13 & 2.10  & 0.46  & 0.5   & \cite{2016MNRAS.461..202A} \\
    TeV J1443-391 & PKS 1440-389 & 0.065 & B   & 16.33 &   --    & 3.10E-11 & 4.75E-12 & 3.38  & 0.27  & 0.274 & 3.63E-11 & 4.97E-12 & 3.05  & 0.25  & 0.274 & \cite{2020MNRAS.494.5590A} \\
    TeV J1512-091 & PKS 1510-089 & 0.361 & F  & 13.93 & 10.5  & 7.73E-11 & 4.17E-12 & 3.05  & 0.10  & 0.175 & 1.22E-10 & 2.07E-11 & 2.42  & 0.35  & 0.175 & \cite{2018AA...619A.159M} \\
    TeV J1517-243 & AP Librae & 0.048 & B   & 14.12 & 20.9  & 4.00E-12 & 4.96E-13 & 2.73  & 0.18  & 0.45  & 5.02E-12 & 6.07E-13 & 2.47  & 0.18  & 0.45  & \cite{2015AA...573A..31H} \\
    TeV J1555+111 & PG 1553+113 & 0.36  & B   & 16.33 & 11.4  & 5.19E-11 & 6.38E-12 & 4.22  & 0.30  & 0.3   & 1.77E-10 & 1.60E-11 & 2.55  & 0.23  & 0.3   & \cite{2015ApJ...799....7A} \\
    TeV J1653+397 & Mkn 501 & 0.034 & B   & 16.81 & 2.3   & 1.74E-10 & 1.04E-11 & 2.48  & 0.06  & 0.3   & 1.86E-10 & 1.09E-11 & 2.28  & 0.06  & 0.3   & \cite{2017AA...603A..31A} \\
    TeV J1725+118 & H 1722+119 & 0.4   & B   & 16.35 & 14.3  & 4.21E-11 & 8.10E-12 & 3.56  & 0.24  & 0.2   & 8.24E-11 & 2.96E-11 & 2.83  & 0.51  & 0.2   & \cite{2016MNRAS.459.3271A} \\
    TeV J1743+196 & 1ES 1741+196 & 0.084 & B   & 15.54 & 17.3  & 9.28E-12 & 2.39E-13 & 2.70  & 0.10  & 0.3   & 1.12E-11 & 4.43E-13 & 2.36  & 0.15  & 0.3   & \cite{2016MNRAS.459.2550A} \\
    TeV J1943+213 & HESS J1943+213 & 0.16  & U   & 16.66 &   --    & 9.61E-12 & 5.81E-13 & 2.81  & 0.10  & 0.4   & 3.24E-11 & 2.82E-12 & 1.81  & 0.13  & 0.3   & \cite{2018ApJ...862...41A} \\
    TeV J1959+651 & 1ES 1959+650 & 0.047 & B   & 17.32 &   --    & 2.67E-12 & 2.28E-13 & 2.54  & 0.11  & 1     & 4.13E-12 & 3.80E-13 & 2.25  & 0.12  & 1     & \cite{2008ApJ...679.1029T} \\
    TeV J2009-488 & PKS 2005-489 & 0.071 & B   & 16.63 & 6.7   & 1.34E-11 & 1.78E-12 & 3.20  & 0.17  & 0.4   & 1.70E-11 & 2.14E-12 & 2.71  & 0.16  & 0.4   & \cite{2010AA...511A..52H} \\
    TeV J2039+523 & 1ES 2037+521 & 0.053 & B   & 15.82 & 53    & 6.40E-12 & 1.13E-13 & 2.27  & 0.02  & 0.4   & 8.06E-12 & 1.37E-13 & 1.99  & 0.02  & 0.4   & \cite{2020ApJS..247...16A} \\
    TeV J2158-302 & PKS 2155-304 & 0.116 & B   & 16.25 & 11.1  & 1.83E-12 & 1.05E-13 & 3.53  & 0.05  & 1     & 5.30E-12 & 3.86E-13 & 2.88  & 0.06  & 1     & \cite{2010AA...520A..83H} \\
    TeV J2250+384 & B3 2247+381 & 0.1187 & B   & 16.55 & 25.7  & 1.36E-11 & 1.17E-12 & 3.31  & 0.13  & 0.3   & 1.94E-11 & 1.79E-12 & 2.69  & 0.14  & 0.3   & \cite{2012AA...539A.118A} \\
    TeV J2347+517 & 1ES 2344+514 & 0.044 & B   & 16.62 & 2.7   & 2.64E-12 & 1.77E-13 & 2.49  & 0.08  & 0.91  & 3.88E-12 & 2.40E-13 & 2.20  & 0.07  & 0.91  & \cite{2017MNRAS.471.2117A} \\
    TeV J2359-306 & H 2356-309 & 0.165 & B   & 16.67 & 14.3  & 3.50E-13 & 4.55E-14 & 2.97  & 0.15  & 1     & 1.89E-12 & 2.03E-13 & 1.89  & 0.12  & 1     & \cite{2010AA...516A..56H} \\
    4FGL J2042.1+2427 & RGB J2042.1+2426 & 0.104 & B   & 16.74 & 15.2  & 2.61E-12 & 3.00E-13 & 2.31  & 0.13  & 0.379 & 4.11E-12 & 6.31E-13 & 1.76  & 0.18  & 0.379 & \cite{2020ApJS..247...16A} \\
\enddata
\tablecomments{
Col. (1): TeV name;
Col. (2): other name; 
Col. (3): redshift; 
Col. (4): classification, B for BLL, F for FSRQ and U for BCU;
Col. (5): synchrotron peak frequency, TeV J0847+115 is from \cite{Fan2016} and the others are from \cite{Yang2022a};
Col. (6): Doppler factor from \cite{Chen2018}; 
Col. (7): the absorbed normalization in units of $\rm cm^{-2} \cdot s^{-1} \cdot TeV^{-1}$; 
Col. (8): the error of the absorbed normalization in units of $\rm cm^{-2} \cdot s^{-1} \cdot TeV^{-1}$; 
Col. (9): the absorbed photon index; 
Col. (10): the error of the absorbed photon index;
Col. (11): the normalization energy in units of TeV; 
Col. (12): the de-absorbed normalization in units of $\rm cm^{-2} \cdot s^{-1} \cdot TeV^{-1}$; 
Col. (13): the error of the de-absorbed normalization in units of $\rm cm^{-2} \cdot s^{-1} \cdot TeV^{-1}$; 
Col. (14): the de-absorbed photon index; 
Col. (15): the error of the de-absorbed photon index; 
Col. (16): the normalization energy in units of TeV; 
Col. (17): the reference of TeV spectra. 
The symbol `--' indicates a null value. 
}
\end{deluxetable*}

\begin{deluxetable*}{lcccccccccccccccc}
\tabletypesize{\tiny}
\tablewidth{0pt} 
% %\tablenum{1}
\tablecaption{The sample of 21 TeV blazars in the flaring state. \label{tab:flaring}}
\rotate 
\tablehead{ 
 TeV name & Other name & z & Class & $\log \, \nu_{p}^{\rm ob}$ & $\delta$ & $N_{0, \rm abs}$   & $\Delta N_{0, \rm abs}$ & $\Gamma_{\rm abs}$ & $\Delta \Gamma_{\rm abs}$ & $E_{0}$ & $N_{0}$ & $\Delta N_{0}$ & $\Gamma$ & $\Delta \Gamma$ & $E_{0}$ & Ref.   \\
}
\colnumbers
\startdata
    TeV J0112+227 & S2 0109+22 & 0.265 & B     & 13.90 & 53.4  & 1.62E-09 & 5.92E-11 & 3.57  & 0.09  & 0.11  & 1.38E-09 & 7.38E-12 & 3.28  & 0.01  & 0.12  & \cite{2018MNRAS.480..879M} \\
    TeV J0218+359 & S3 0218+35 & 0.944 & F     & 12.99 &   --    & 2.18E-09 & 2.17E-10 & 3.91  & 0.45  & 0.1   & 4.28E-09 & 7.75E-10 & 2.45  & 0.89  & 0.1   & \cite{2016AA...595A..98A} \\
    TeV J0521+211 & VER J0521+211 & 0.108 & B     & 15.24 & 14.3  & 2.35E-10 & 1.96E-11 & 3.16  & 0.12  & 0.3   & 3.32E-10 & 1.44E-11 & 2.72  & 0.07  & 0.3   & \cite{2022ApJ...932..129A} \\
    TeV J0721+713 & S5 0716+714 & 0.31  & B     & 14.42 & 20.3  & 1.23E-09 & 7.35E-11 & 4.65  & 0.16  & 0.15  & 1.47E-09 & 1.27E-10 & 3.65  & 0.24  & 0.15  & \cite{2018AA...619A..45M} \\
    TeV J0809+523 & 1ES 0806+524 & 0.138 & B     & 16.26 & 3.7   & 1.27E-11 & 1.65E-12 & 2.92  & 0.19  & 0.5   & 2.75E-11 & 2.93E-12 & 2.19  & 0.16  & 0.5   & \cite{2015MNRAS.451..739A} \\
    TeV J0854+201 & OJ 287 & 0.3056 & B     & 13.62 & 67.5  & 1.48E-11 & 1.15E-12 & 2.18  & 0.11  & 0.1   & 1.59E-11 & 3.76E-12 & 1.42  & 0.33  & 0.1   & \cite{2009PASJ...61.1011S} \\
    TeV J0958+655 & S4 0954+65 & 0.367 & B     & 14.04 & 26.5  & 1.01E-09 & 5.63E-11 & 4.64  & 0.24  & 0.15  & 1.38E-09 & 2.86E-11 & 3.98  & 0.09  & 0.15  & \cite{2018AA...617A..30M} \\
    TeV J1104+382 & Mkn 421 & 0.031 & B     & 16.61 & 1.5   & 2.49E-11 & 1.24E-12 & 2.28  & 0.06  & 1     & 3.31E-11 & 1.71E-12 & 2.09  & 0.06  & 1     & \cite{2009ApJ...703..169A} \\
    TeV J1159+292 & Ton 599 & 0.7247 & F     & 13.54 & 29.3  & 3.08E-10 & 1.76E-11 & 4.37  & 0.13  & 0.2   & 1.56E-09 & 3.15E-10 & 0.16  & 0.61  & 0.2   & \cite{2022ApJ...924...95A} \\
    TeV J1217+301 & 1ES 1215+303 & 0.131 & B     & 15.35 & 15.1  & 9.42E-11 & 6.82E-12 & 3.59  & 0.14  & 0.4   & 8.49E-11 & 6.77E-12 & 2.93  & 0.12  & 0.5   & \cite{2020ApJ...891..170V} \\
    TeV J1221+282 & W Comae & 0.103 & B     & 15.54 & 13.8  & 2.11E-11 & 3.23E-12 & 3.72  & 0.37  & 0.4   & 3.19E-11 & 4.39E-12 & 3.11  & 0.35  & 0.4   & \cite{2008ApJ...684L..73A} \\
    TeV J1224+213 & 4C +21.35 & 0.435 & F     & 14.06 & 11.3  & 7.85E-10 & 9.37E-11 & 3.73  & 0.21  & 0.2   & 1.85E-09 & 5.69E-11 & 2.63  & 0.06  & 0.2   & \cite{2011ApJ...730L...8A} \\
    TeV J1256-057 & 3C 279 & 0.5362 & F     & 12.79 & 27.7  & 4.45E-10 & 1.08E-10 & 4.33  & 0.44  & 0.2   & 8.66E-10 & 5.20E-10 & 3.75  & 1.03  & 0.2   & \cite{2008Sci...320.1752M} \\
    TeV J1422+323 & B2 1420+32 & 0.682 & F     & 13.63 & 4.3   & 2.41E-09 & 2.35E-10 & 4.12  & 0.18  & 0.1   & 3.75E-09 & 5.15E-10 & 3.49  & 0.27  & 0.1   & \cite{2021AA...647A.163M} \\
    TeV J1443+250 & PKS 1441+25 & 0.93978 & F     & 13.99 &   --    & 1.13E-09 & 1.14E-10 & 4.63  & 0.24  & 0.1   & 2.59E-09 & 1.25E-10 & 3.16  & 0.14  & 0.1   & \cite{2015ApJ...815L..23A} \\
    TeV J1512-091 & PKS 1510-089 & 0.361 & F     & 13.93 & 10.5  & 3.27E-10 & 5.10E-11 & 4.22  & 0.18  & 0.265 & 8.46E-10 & 8.63E-11 & 3.34  & 0.13  & 0.265 & \cite{2021AA...648A..23H} \\
    TeV J1653+397 & Mn 501 & 0.034 & B     & 16.81 & 2.3   & 6.32E-10 & 4.56E-11 & 2.26  & 0.07  & 0.3   & 6.76E-10 & 4.95E-11 & 2.06  & 0.07  & 0.3   & \cite{2017AA...603A..31A} \\
    TeV J1728+502 & 1ES 1727+502 & 0.055 & B     & 16.52 & 2     & 7.71E-12 & 9.10E-13 & 2.19  & 0.24  & 0.62  & 1.09E-11 & 1.30E-12 & 1.82  & 0.24  & 0.62  & \cite{2015ApJ...808..110A} \\
    TeV J2202+422 & BL Lacertae & 0.069 & B     & 13.93 & 3.8   & 5.55E-10 & 7.96E-11 & 3.69  & 0.48  & 0.3   & 6.50E-10 & 9.07E-11 & 3.38  & 0.47  & 0.3   & \cite{2013ApJ...762...92A} \\
    TeV J2243+203 & RGB J2243+203 & 0.39  & B     & 15.14 & 14.3  & 6.33E-10 & 4.79E-11 & 4.62  & 0.24  & 0.15  & 8.69E-10 & 1.05E-10 & 3.44  & 0.44  & 0.15  & \cite{2017ApJS..233....7A} \\
    TeV J2347+517 & 1ES 2344+514 & 0.044 & B     & 16.62 & 2.7   & 1.73E-11 & 1.27E-12 & 2.44  & 0.14  & 1     & 2.61E-11 & 1.83E-12 & 2.14  & 0.13  & 1     & \cite{2011ApJ...738..169A} \\
\enddata
\tablecomments{
Col. (1): TeV name;
Col. (2): other name; 
Col. (3): redshift; 
Col. (4): classification, B for BLL and F for FSRQ;
Col. (5): synchrotron peak frequency from \cite{Yang2022a}; 
Col. (6): Doppler factor from \cite{Chen2018}; 
Col. (7): the absorbed normalization in units of $\rm cm^{-2} \cdot s^{-1} \cdot TeV^{-1}$; 
Col. (8): the error of the absorbed normalization in units of $\rm cm^{-2} \cdot s^{-1} \cdot TeV^{-1}$; 
Col. (9): the absorbed photon index; 
Col. (10): the error of the absorbed photon index;
Col. (11): the normalization energy in units of TeV; 
Col. (12): the de-absorbed normalization in units of $\rm cm^{-2} \cdot s^{-1} \cdot TeV^{-1}$; 
Col. (13): the error of the de-absorbed normalization in units of $\rm cm^{-2} \cdot s^{-1} \cdot TeV^{-1}$; 
Col. (14): the de-absorbed photon index; 
Col. (15): the error of the de-absorbed photon index; 
Col. (16): the normalization energy in units of TeV; 
Col. (17): the reference of TeV spectra. 
The symbol `--' indicates a null value. 
}
\end{deluxetable*}

% \section{The fitting normalizations and parameters of each source for the \textit{Fermi}-LAT }

\begin{deluxetable}{lcccccccl}
\tabletypesize{\footnotesize}
\tablewidth{0pt} 
% %\tablenum{1}
\tablecaption{The normalizations and parameters of spectrum in the quiescent state. \\ \label{tab:fermi_quiescent}}
%\rotate 
\tablehead{ 
TeV name & $N_0$  & $\Delta N_0$  & $\Gamma$ & $\Delta \Gamma$\ & $\beta$  & $\Delta \beta$ & $E_0$ & Observation periods\\
}
\colnumbers
\startdata
    TeV J0013-188$^{\dag}$ & 7.6E-14 & 3.18E-14 & 1.71  & 0.26  & --    & --    & 2498.28394 & MJD 54683 - 55912 \\
    TeV J0033-193 & 1.36E-12 & 1.77E-13 & 1.74  & 0.11  & 0.10  & 0.06  & 1571.1472 & MJD 55145 - 56954 \\
    TeV J0112+227 & 3.29E-11 & 7.19E-12 & 1.71  & 0.26  & 0.06  & 0.10  & 755.25354 & MJD 57225 - 57231 \\
    TeV J0214+517$^{\dag}$ & 3.71E-14 & 1.75E-14 & 1.66  & 0.32  & --    & --    & 4042.61401 & MJD 57370 - 58042 \\
    TeV J0303-241 & 4.35E-12 & 7.08E-13 & 2.06  & 0.13  & 0.01  & 0.06  & 945.1168 & MJD 55044 - 55895 \\
    TeV J0319+187$^{\dag}$ & 1.51E-14 & 8.52E-15 & 1.43  & 0.31  & --    & --    & 6216.29443 & MJD 54732 - 55485 \\
    TeV J0416+010$^{\dag}$ & 4.16E-14 & 5.19E-15 & 1.67  & 0.07  & --    & --    & 3730.71851 & MJD 54683 - 55158 \\
    TeV J0449-438 & 4.52E-12 & 4.03E-13 & 1.87  & 0.08  & 0.03  & 0.04  & 1581.183 & MJD 55136 - 55227 \\
    TeV J0509+056 & 2.04E-11 & 1.46E-12 & 2.09  & 0.05  & 0.06  & 0.03  & 1074.2167 & MJD 58019.39 - 58155.18 \\
    TeV J0521+211 & 5.43E-12 & 9.49E-15 & 1.81  & 1.23E-03  & 0.01  & 4.33E-04  & 1542.1824 & MJD 55126 - 55212 \\
    TeV J0648+152$^{\dag}$ & 1.54E-14 & 3.87E-15 & 2.92  & 0.09  & --    & --    & 4960.08594 & MJD 55259 - 55301 \\
    TeV J0710+591$^{\dag}$ & 3.68E-14 & 1.4E-14 & 1.50  & 0.23  & --    & --    & 4958.52441 & MJD 54801 - 54921 \\
    TeV J0809+523 & 1.88E-12 & 3.46E-13 & 2.27  & 0.14  & 2.13E-06 & 1.16E-03 & 1345.8887 & MJD 55568 - 55622 \\
    TeV J1217+301 & 6.85E-12 & 6.52E-14 & 1.82  & 0.01  & 0.06  & 3.28E-03  & 1093.8949 & MJD 57202 - 57562 \\
    TeV J1221+301$^{\dag}$ & 2.09E-13 & 3.53E-14 & 1.54  & 0.11  & --    & --    & 4501.46387 & MJD 54801 - 54952 \\
    TeV J1422+323 & 3.96E-10 & 1.24E-12 & 1.87  & 2.63E-03 & 0.04  & 1.08E-03 & 579.01245 & MJD 58873.5 - 58880.5 \\
    TeV J1427+238 & 7.55E-12 & 5.36E-13 & 1.76  & 0.06  & 0.04  & 0.02  & 1204.567 & MJD 54881 - 55003 \\
    TeV J1443+120$^{\dag}$ & 6.28E-14 & 1.68E-14 & 1.79  & 0.20  & --    & --    & 3828.12671 & MJD 54683 - 55362 \\
    TeV J1443-391 & 7.18E-13 & 1.68E-13 & 1.25  & 0.15  & 0.20  & 0.07  & 1978.045 & MJD 55985 - 56074 \\
    TeV J1512-091 & 7.26E-11 & 2.27E-12 & 2.45  & 0.02  & 0.02  & 0.01  & 723.6382 & MJD 56000 - 58000 \\
    TeV J1517-243 & 8.55E-12 & 1.15E-12 & 2.32  & 0.12  & 6.77E-09 & 3.46E-05 & 811.1857 & MJD 55326 - 55689 \\
    TeV J1555+111 & 4.86E-12 & 3.36E-13 & 1.66  & 0.05  & 0.07  & 0.03  & 1802.4828 & MJD 55317 - 56108 \\
    TeV J1653+397 & 4.73E-12 & 4.49E-14 & 1.71  & 0.01  & 2.00E-10 & 5.10E-07 & 1486.4363 & MJD 54907 - 55004 \\
    TeV J1743+196$^{\dag}$ & 2.7E-14 & 2.54E-14 & 1.25  & 0.37  & --    & --    & 2791.80957 & MJD 54940 - 56834 \\
    TeV J2347+517 & 2.65E-13 & 2.21E-14 & 1.70  & 0.04  & 0.01  & 0.01  & 1934.7299 & MJD 54683 - 57023 \\
\enddata
\tablecomments{$\dag$: power-law function. The other sources are log-parabola function. \\
Col. (1): TeV name; 
Col. (2):the normalization in units of $\rm cm^{-2} \cdot s^{-1} \cdot MeV^{-1}$; 
Col. (3): the error of normalization in units of $\rm cm^{-2} \cdot s^{-1} \cdot MeV^{-1}$; 
Col. (4): the photon index; 
Col. (5): the error of the photon index; 
Col. (6): the curvature index of log-parabola function; 
Col. (7): the error of the curvature index; 
Col. (8): the pivot energy in units of MeV; 
Col. (9): the same observation periods as the VHE observation.
The symbol `--' indicates a null value. 
}
\end{deluxetable}

\begin{deluxetable}{lcccccccl}
\tabletypesize{\footnotesize}
\tablewidth{0pt} 
% %\tablenum{1}
\tablecaption{The normalizations and parameters of spectrum in the flaring state.  \label{tab:fermi_flaring}}
% \rotate 
\tablehead{ 
TeV name & $N_0$  & $\Delta N_0$  & $\Gamma$ & $\Delta \Gamma$ & $\beta$  & $\Delta \beta$ & $E_0$ & Observation periods \\
}
\colnumbers 
\startdata
    TeV J0112+227 & 6.27E-11 & 2.67E-11 & 1.73  & 0.38  & 0.16  & 0.21  & 755.25354 & MJD 57228 \\
    TeV J0218+359 & 5.86E-11 & 1.49E-11 & 1.30  & 0.23  & 0.30  & 0.13  & 764.40204 & MJD 56863 \\
    TeV J0521+211 & 3.05E-11 & 7.51E-12 & 1.90  & 0.22  & 0.31  & 0.20  & 1542.1824 & MJD 56628.5 - 56632.5 \\
    TeV J0721+713 & 1.34E-10 & 9.44E-12 & 1.86  & 0.05  & 0.03  & 0.03  & 734.5862 & MJD 57040 - 57050 \\
    TeV J0809+523 & 1.39E-11 & 8.02E-12 & 2.14  & 0.49  & 0.14  & 0.30  & 1345.8887 & MJD 55616 \\
    TeV J0958+655 & 3.94E-10 & 5.94E-11 & 1.83  & 0.14  & 0.11  & 0.08  & 651.3296 & MJD 57067 \\
    TeV J1159+292 & 6.2E-10 & 6.11E-11 & 1.65  & 0.11  & 0.05  & 0.04  & 523.50714 & MJD 58102 - 58104 \\
    TeV J1217+301 & 5.74E-11 & 2.13E-11 & 1.09  & 0.48  & 0.17  & 0.15  & 1093.8949 & MJD 57844 \\
    TeV J1224+213 & 2.93E-09 & 2.49E-10 & 1.70  & 0.09  & 0.08  & 0.03  & 392.393 & MJD 55364 \\
    TeV J1422+323$^{\dag}$ & 3.28E-10 & 2.66E-11 & 1.90  & 0.06  & --    & --    & 749.958679 & MJD 58868.3 - 58870.3  \\
    TeV J1443+250 & 8.28E-11 & 8.22E-12 & 1.60  & 0.10  & 0.11  & 0.04  & 850.17804 & MJD 57130 - 57139.5 \\
    TeV J1512-091 & 2.57E-10 & 4.75E-11 & 1.74  & 0.16  & 0.01  & 0.06  & 723.6382 & MJD 57538 \\
    TeV J1653+397 & 9.38E-12 & 3.96E-12 & 1.48  & 0.27  & 0.05  & 0.10  & 1486.4363 & MJD 54952.41 - 54955 \\
    TeV J2202+422 & 1.63E-10 & 4.12E-11 & 1.92  & 0.24  & 0.33  & 0.21  & 796.1543 & MJD 55739.5 - 55740.5 \\
    TeV J2243+203 & 1.01E-11 & 5.48E-12 & 1.74  & 0.37  & 9.64E-09 & 3.98E-05 & 1517.4087 & MJD 57012 \\
\enddata
\tablecomments{$\dag$: power-law function. The other sources are log-parabola function. \\
Col. (1): TeV name; 
Col. (2):the normalization in units of $\rm cm^{-2} \cdot s^{-1} \cdot MeV^{-1}$;
Col. (3): the error of normalization in units of $\rm cm^{-2} \cdot s^{-1} \cdot MeV^{-1}$; 
Col. (4): the photon index; 
Col. (5): the error of the photon index; 
Col. (6): the curvature index of log-parabola function; 
Col. (7): the error of the curvature index; 
Col. (8): the pivot energy in units of MeV; 
Col. (9): the same observation periods as the VHE observation.
The symbol `--' indicates a null value. 
}
\end{deluxetable}

\appendix 

\section{The fitting result of VHE spectrum}
\begin{figure}[htbp]
    \centering
    \begin{subfigure}{\textwidth}
    \includegraphics[width=7 in]{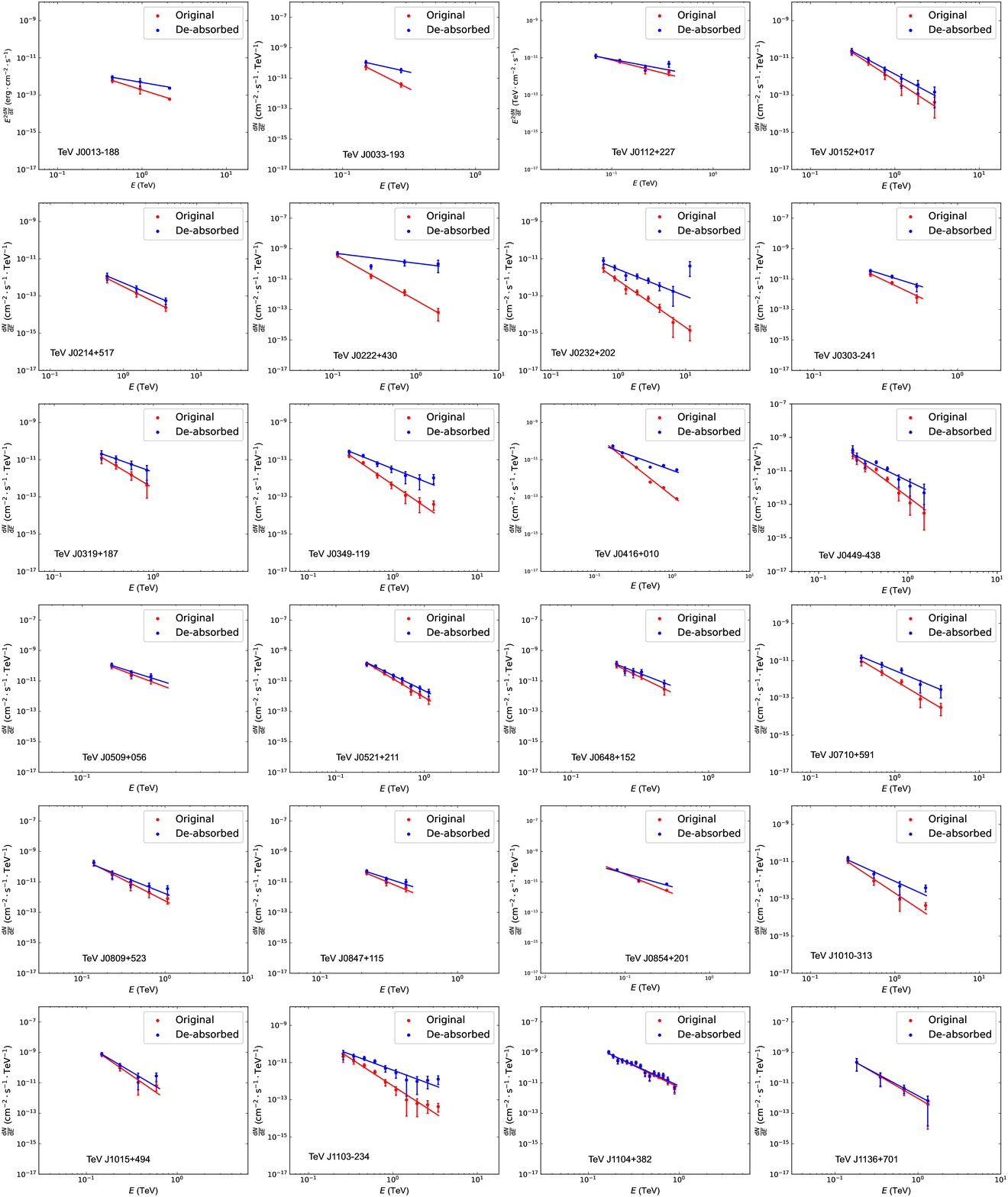}
    \subcaption{}
    \label{a}
    \end{subfigure}
    % \label{fig:VHE_q1}
\end{figure}

\begin{figure}[htbp]
    \ContinuedFloat
    \centering
    \begin{subfigure}{\textwidth}
    \includegraphics[width=7 in]{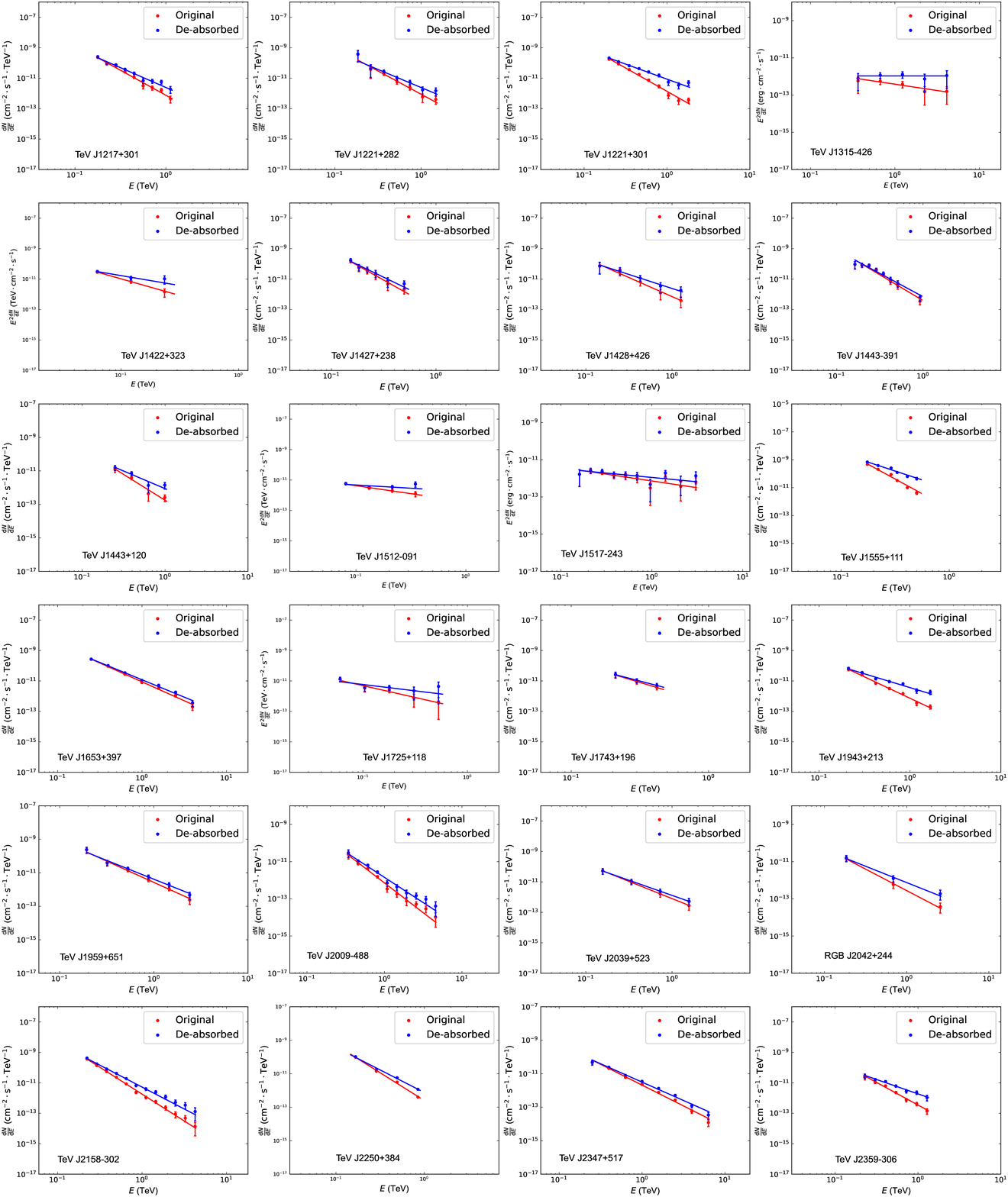}
    \subcaption{}
    \label{b}
    \end{subfigure}
    \caption{The fitting VHE spectrum of the quiescent state for each source. The red circle for the original spectrum and the blue one for the de-absorbed spectrum corrected by the EBL model from \citet{Franceschini2008}. } 
    \label{fig:VHE_q}
\end{figure}

\begin{figure}[htbp]
    \centering
    \includegraphics[width=7 in]{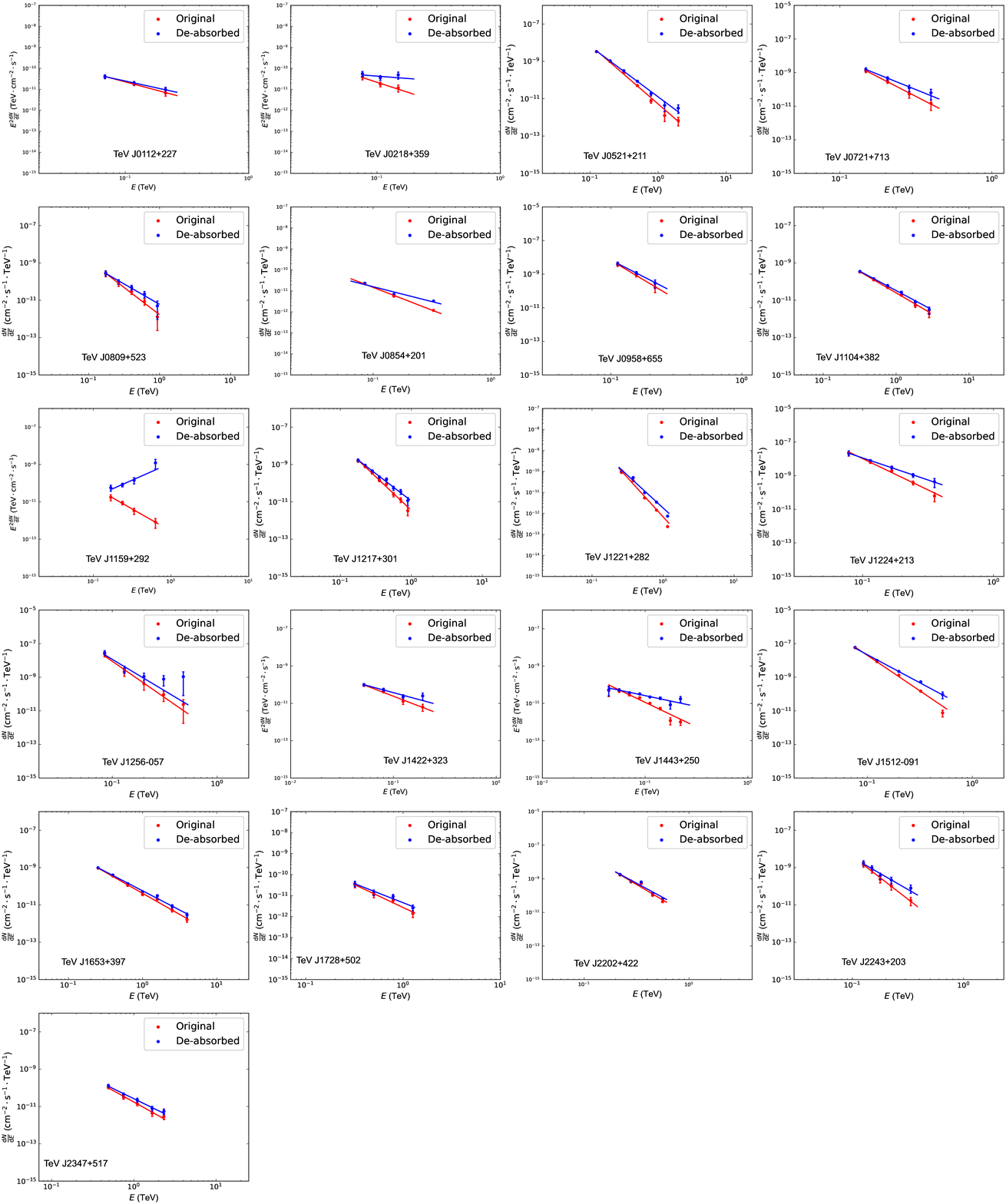}
    \caption{The fitting VHE spectrum of the flaring state for each source. The red circle for the original spectrum and the blue one for the de-absorbed spectrum corrected by the EBL model from \citet{Franceschini2008}. }
    \label{fig:VHE_f}
\end{figure}

\end{document}